\documentclass[a4paper,fleqn]{cas-dc}
\usepackage{placeins}

\usepackage[sort,numbers]{natbib}

\def\tsc#1{\csdef{#1}{\textsc{\lowercase{#1}}\xspace}}
\tsc{WGM}
\tsc{QE}
\tsc{EP}
\tsc{PMS}
\tsc{BEC}
\tsc{DE}
\usepackage{graphicx}
\usepackage{lscape}
\usepackage{array}
\usepackage{graphicx}
\usepackage{multicol}
\usepackage{multirow}
\usepackage{amssymb}
\usepackage{pifont}
\usepackage{tikz}
\usepackage{verbatim}
\usepackage{array}
\usepackage{longtable}
\usepackage{supertabular}

\newcommand{\xmark}{\text{\ding{55}}}
\begin{document}
\let\WriteBookmarks\relax
\def\floatpagepagefraction{1}
\def\textpagefraction{.001}
\shorttitle{Education Through Metaverses}
\shortauthors{Senthil Kumar J et~al.}

\title [mode = title]{Advancing Education Through Extended Reality and Internet of Everything Enabled Metaverses: Applications, Challenges, and Open Issues}                      



\author[1]{Senthil Kumar Jagatheesaperumal}[
                        orcid=0000-0002-9516-0327]
\ead{senthilkumarj@mepcoeng.ac.in}


\address[1]{Department of Electronics and Communication Engineering, Mepco Schlenk Engineering College, Sivakasi, Tamil Nadu, India}

\author[2]{ Kashif Ahmad}[type=editor,
                          orcid=0000-0002-0931-9275]
\cormark[1]
\ead{kashif.ahmad@mtu.ie}
\address[2]{Department of Computer Science, Munster Technological University, Cork, Ireland.}

\author[3]{ Ala Al-Fuqaha}[%
   orcid=0000-0002-0903-1204]
\ead{aalfuqaha@hbku.edu.qa}

\address[3]{Information and Computing Technology (ICT) Division, College of Science and Engineering (CSE), Hamad Bin Khalifa University, Doha, Qatar.}

\author[4]{ Junaid Qadir}[%
    orcid=0000-0001-9466-2475]
\ead{jqadir@qu.edu.qa}



\address[4]{Department of Computer Science and Engineering, College of Engineering, Qatar University, Doha, Qatar.}

\cortext[cor2]{Corresponding Author Name: Ala Al-Fuqaha; Department of Computer Science, Munster Technological University, Cork, Ireland; Email ID: aalfuqaha@hbku.edu.qa}


\begin{abstract}
Metaverse has evolved as one of the popular research agendas that let the users learn, socialize, and collaborate in a networked 3D immersive virtual world. Due to the rich multimedia streaming capability and immersive user experience with high-speed communication, the metaverse is an ideal model for education, training, and skill development tasks. To facilitate research in this area, we provide a comprehensive review of the various educational use cases and explore how enabling technologies such as Extended reality (XR) and Internet of Everything (IoE) will play a major role in educational services in future metaverses. Secondly, we provide an overview of metaverse-based educational applications focusing on education, training, and skill development and analyze the technologies they are built upon. We identify common research problems and future research directions in the domain. The paper also identifies core ethical considerations of metaverse for education and potential pitfalls. We believe this survey can fully demonstrate the versatility of metaverse-driven education, which could serve as a potential guideline for the researchers. 

\end{abstract}




\maketitle
\section{Introduction}
\label{introduction}
In the modern world, the popularity of the metaverse is on the rise in different application domains. Some key application domains where the technology has been proved very effective include healthcare, defense, industry (manufacturing), real-estate, and gaming \cite{ning2021survey}. The vision of metaverse is driven by advances in technologies such as artificial intelligence (AI), extended reality (XR), and internet of everything (IoE).

Education is also one of the domains where the use of the concept/technology is getting momentum with metaverse promising several advantages. For instance, it allows students and teachers from different parts of the world to meet up in a virtual environment regardless of their real-world location \cite{metaversemeetup}. Similarly, building virtual landscapes on the basis of the teacher's lesson plans provides better opportunity, resulting in an improved and more productive learning experience for the students. Due to such opportunities and the advantages, it brings to the education sector, several studies have analyzed the potential of a metaverse in education \cite{park2022metaverse} as described next. 

The potential of a metaverse in education was explored by Sung et al. \cite{sung2021development}, who compared it with traditional educational content delivery based on video presentations. The evaluation is carried out in terms of students' learning attitude, enjoyment, and performance in a knowledge-based test. It helps to revitalize the economy and market through XR technology. Similarly, Kemp et al. \cite{kemp2006putting} analyzed the pros and cons of a multi-user virtual environment in the education sector. Here, the integration of traditional learning with the multi-user virtual environments and a hybrid educational platform was established. It encompasses interactive objects and learning spaces. Therefore, envisioned to support a typical diversified range of educational services with immersive experiences for the users, metaverse will most likely make ground-breaking innovations through IoE and XR. 

Despite the great potential and proven effectiveness in student engagement in learning, the supporting technologies for metaverse also bring several challenges to the education sector. For example, as demonstrated by Maccallum et al. \cite{maccallum2019teacher}, generating virtual content and new ideas for engaging students in the learning process is very challenging and requires a degree of experience. Some studies also report the little impact of the technology on the students' performance in knowledge-based tests \cite{sung2021development}. 

\subsection{Enabling Technologies for Metaverse}

Metaverse is based on the crossroads of key modern technologies namely XR and IoE. In the next subsections, we discuss these elements of a metaverse in detail by highlighting their roles and the advantages they bring to a metaverse. 

\subsubsection{Extended Reality (XR)} 
\begin{figure}[!ht]
  \centering    
  \includegraphics[width=0.5\textwidth]{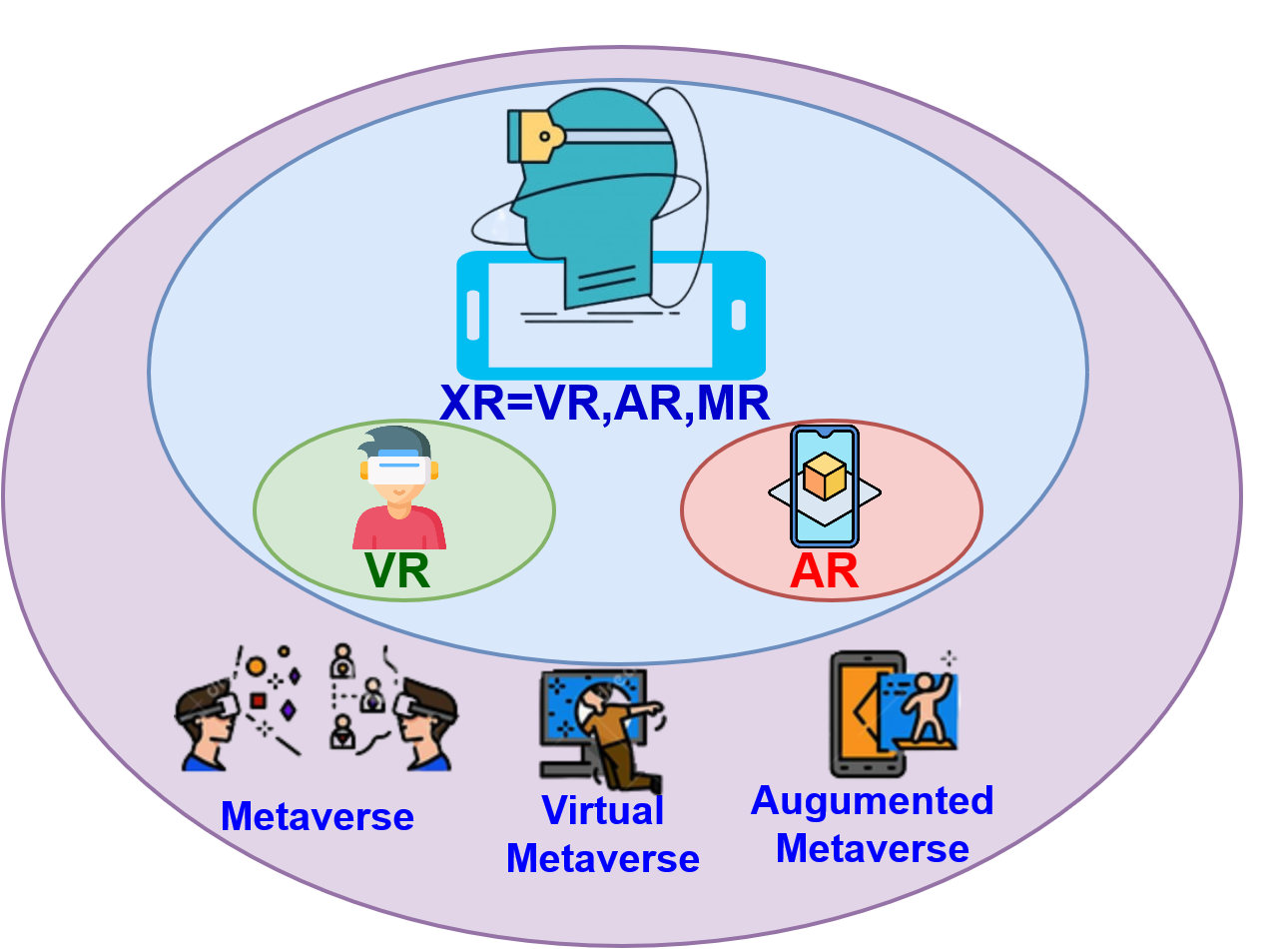}
   \caption{The Digital World Realization under the umbrella of XR}
   \label{fig:vrarxr}
\end{figure}

XR drags and stretches the human experiences by blending the real and virtual digital worlds in multi-dimensional directions. As shown in Fig.~\ref{fig:vrarxr}, XR is an umbrella that integrates VR, AR and MR technology for provisioning sustainable digital world realization.

\begin{itemize}
    \item \textit{VR:} In general, VR creates a whole new environment and provides a completely immersive experience for the users. It uses computer technology to create a simulated experience, that may be similar or completely different from the real world. Standard VR systems use either headsets or multi-projected environments to generate realistic sounds and visuals.
    \item \textit{AR:}  AR, on the other hand, keeps the real-world objects as it is and superimposes layers of digital objects to the real world. AR systems integrate three different features which are 1) the combination of the real and virtual world, 2) a real-time interaction, and 3) accurate 3D registration of virtual and real objects. The following are the most vital components used for providing rich AR experiences for end-users.
    \item \textit{MR:} MR is the merging of real and virtual worlds to produce new environments and visualizations. Here, the physical and digital objects coexist and interact in real-time~\cite{bec2021virtual}. Unlike AR, users can interact with virtual objects. To provide different user experiences from fully immersive to light information layering of environments, MR developers have provided robust tools to bring up virtual experiences to life.
\end{itemize}
VR and AR are the key components of realizing the vision of a 3D immersive metaverse experience. MR relies on the interaction between physical and virtual world objects.  These technologies help in truly owning a digital space allowing for innovation and creativity without any central authority \cite{sparkes2021metaverse}. As XR are at present being a core part of the Metaverse, the holograms are supposed to be the next big counterpart that could assist the Metaverse. XR imparts natural technological progression and it allows people to be independent of being anywhere in the world. Thanks to the recent advancement in technology, AR applications can also be run on mobile devices ensuring interaction with virtual and real-world objects. This paves the way for Mobile Augmented Reality (MAR) in the metaverse~\cite{siriwardhana2021survey}. 


\subsubsection{Internet of Everything (IoE)}
The IoE infrastructure is built on top of smart things, data, processes and people. However, the IoT systems are built around smart things alone. IoE plays a key role in delivering rich metaverse applications.The following technologies contribute towards successful incorporation of IoE infrastructure.

\begin{itemize}
    \item \textit{Web 3.0:} Web 3.0, which is the next evolution of the World Wide Web, drives the next version of the Internet. It helps to own parts of the Internet, unlike web 2.0 which provides the owner of the domain name only. The applications and services of Web 3.0 are mainly powered by distributed ledger technology. Web 3.0 makes us a legitimate owner of our content by providing complete authority over the content and forces every organization to operate in a decentralized manner. Decentralized Autonomous Organizations (DAO), will not have CEOs or presidents for those organizations, instead, shareholders share tokens and make changes through voting. Web 3.0 may not connect the digital identity of a person completely with the real identity of the person, which means digital transactions made by that person will not reveal their real identity. Web 3.0 and the metaverse complement each other \cite{metaverseweb3}. Web 3.0 provides connectivity services in a metaverse while the metaverse provides the basis for the decentralized implementation of Web 3.0 services. 
    
    \item \textit{5G/6G Services:} The last mile latency normally encountered in other communication services is significantly reduced to a larger extent with the advancements in 5G/6G services. Further, its impact on users' experiences in metaverse has improved drastically by eliminating the jitters and provisioning larger bandwidth and speed. It also helps to boost the user experiences by providing multi-access edge computing, and universal as well as standard edge offloading services. With metaverse being the next generation of the Internet, most companies have already sensed the transformation and the demand for adopting 5G and 6G communication services~\cite{mozumder2022overview}. Such high-speed communication services alongside the blockchain and AI for building a digital virtual world are highly supportive of the metaverse. 

\end{itemize}


\subsection{Layers of the Metaverse}
\begin{itemize}

\item \textit{Experience:}
Beyond the workplace and home for most people, Metaverse provides the experience of "third place", with venues for immersive social life, community interaction, shopping, e-sports, and other various activities through creativity. This aspect/layer of the metaverse is one of the main causes of the buzz and the investment attracted it \cite{metaversebuzz}. The experience layer of Metaverse dematerializes the physical space and objects by incorporating social immersive experience. 

\vspace{1mm}
\item \textit{Discovery:}
Instead of focusing on what the people need in the network, this layer visualizes what people are doing at present. The metaverse provides both inbound and outbound discoveries. Some common ways of inbound discoveries include search engines, community-driven content sharing, and real-time prescience \cite{metaversebuzz}. The outbound discoveries in the metaverse occur through notifications, emails, social media posts, and advertisements. Real-time presence and community-driven content sharing are more economic means of marketing and discovery. Here interaction among the people allows achieving a shared experience by making good use of the real-time presence of people and enabling virtual interaction. 

\vspace{1mm}
\item \textit{Creator Economy:}
There is a huge business potential in the virtual world resulting from the metaverse and further, the content creators are expecting to witness significant growth in the economy. More companies are expected to invest in metaverse for building a sustainable economy. In this regard, the content creators, who are already enjoying great success in different social media outlets, are going to play a major role \cite{metaversebuzz}. Creators are already playing their part with the help of integrated full suite tooling, networking, and discovery for monetization of the economy and crafting the experience of the users.

\vspace{1mm}
\item \textit{Spatial Computing:}
The machines and devices involved in providing the Metaverse experience need not be tied together to a fixed location. Further, spatial computing imparts this feature, which is synonymous with XR, and it intends to make a very big leap in the economic transformation of the Metaverse. It also helps to bind the community through shared services and thereby incorporates the feature of digital togetherness.

\vspace{1mm}
\item \textit{Decentralization:}
In the process of decentralization, decision-making authorization control transfers from a centralized entity to a distributed authorized network. It is capable of providing a scalable ecosystem for Metaverse without the need for focusing on the integration of the back-end capabilities of interoperable systems.

\vspace{1mm}
\item \textit{Human Interface:}
This layer of the metaverse is concerned about the hardware devices and technology that allows humans to interact with the machines. A better human interface mechanism will allow users to experience the true potential of the metaverse. Thanks to the recent advance in the technology, several connected handheld devices, 3D-printed wearable, and biosensors are available to bring humans closer to machines. This allows to build more immersive applications for the Metaverse. Furthermore, the integration of embedded AI technology for the metaverse imparts low-latency edge-computing solutions for gaining immersive experiences.

\vspace{1mm}
\item \textit{Infrastructure:}
The infrastructure layer of the metaverse provides the technological infrastructure needed to build a fully functional and interoperable metaverse. The infrastructure of the metaverse is composed of different technologies, such as computational and communication resources, machine intelligence, blockchain technology, gaming, and display technologies. However, improvement in the communication speed, with reduced latency, is in demand for delivering rich content without any network contention and latency. 5G and 6G services could be supported with high speed and better utilization of the bandwidth. Furthermore, the performance enhancement with minimum hardware requirements and power sources are also playing a vital role in building the infrastructure for the Metaverse.

\end{itemize}

\subsection{Scope of the paper}
\begin{figure*}[!t]
\centering
\centerline{\includegraphics[height=12cm, width=18cm]{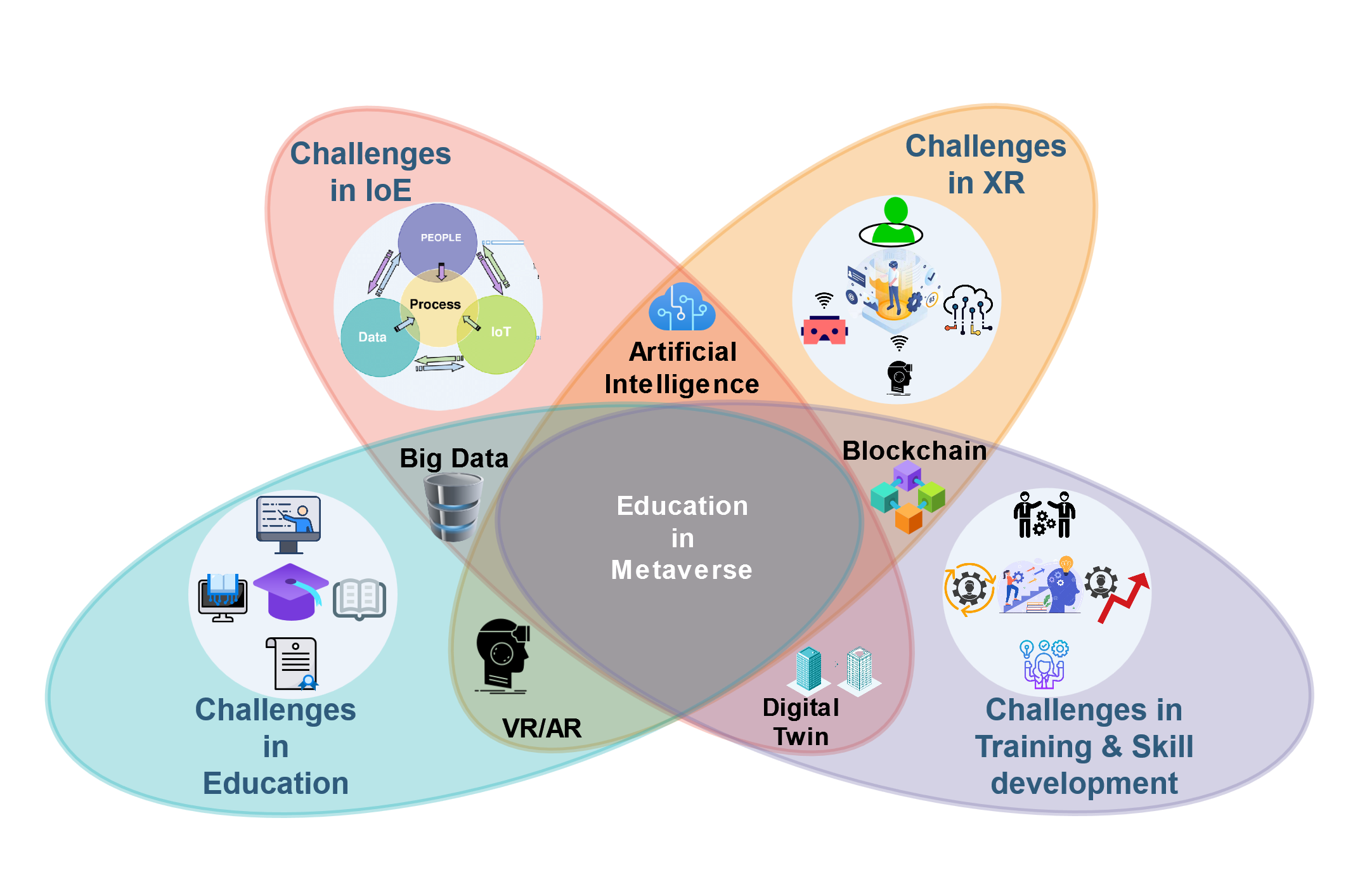}}
\caption{Visual depiction of the scope of the paper.} 
\label{fig:Scope}
\end{figure*}

This paper revolves around the applications of the metaverse and its supporting technologies in the education sector. The paper emphasizes the importance of metaverse and its impact on key applications of education, skill development, and training. We also discuss the related concepts and supporting technologies in detail by analyzing the convergence of XR, and IoE for metaverse educational use cases. 
The paper also highlights the current limitations and drawbacks of the technology in the sector. We also analyze research trends in the domain by identifying the aspects of the technology that need the attention of the research community. Fig.~\ref{fig:Scope} shows the visual description of the scope of the paper.



\subsection{Related Surveys}
The literature reports extensive work on the key enabling technologies such as XR and IoEs. The literature also reports extensive research on the joint applications of XR and IoE, where they are used for providing immersive experiences for diversified range of applications in different domains ~\cite{hu2021virtual, lanka2017review, minerva2020digital}. However, the metaverse is considered to be a different discipline, even though they share the same applications and concerns \cite{cheng2022will}. The literature also reports several interesting works and surveys on the emerging concept of metaverse \cite{lee2021all}. For instance, Ning et al. ~\cite{ning2021survey} survey the literature on metaverse concepts, state-of-the-art techniques, applications, and associated challenges. 
Similarly, Wang et al. \cite{wang2022survey} provided a detailed survey of the literature with a particular focus on the fundamental concepts and key privacy and security-related issues. There are some works \cite{kye2021educational,mozumder2022overview,han2021dynamic} that see XR and IoE as converging subjects for the metaverse applications. In this survey, we focus on the convergence of these technologies for different applications in the education sector. The synthesis of research directions in this paper is handled in a smooth progression, starting from the fundamentals of the metaverse, followed by its technical building platforms, through to its role in provisioning immersive educational services. 
Accordingly, the reader will gradually become familiar with state-of-the-art and supporting techniques such as IoE, and XR whilst gaining an insight into the challenges and opportunities in the broad areas of educational services and skill development through metaverse frameworks. Table \ref{tab:survey} provides a summary of the existing surveys on metaverse.

\subsection{Contribution of this Paper}
The primary objective of this paper is to provide the reader with a comprehensive survey of the literature on metaverse and associated technologies for education, skill development, and training. The contributions of the paper are manifold. On the one side, it provides a complete tutorial on the metaverse by describing the key components and supporting technologies. On the other side, it provides a detailed overview of the existing literature on the topic by highlighting key challenges, potential opportunities, and future research directions. The key contributions of this work can be summarized as follows:


\begin{itemize}
    \item The paper mainly focuses on the convergence of key technologies namely IoE, XR, 
    in the metaverse for education, training, and skill development. 
    


    \item  The paper provides a detailed survey of the state-of-the-art metaverse techniques that leverage XR and IoE paradigms for education, training, and skill development tasks. 

    \item  The paper also elaborates on the description of metaverse approaches for adaptive educational services, such as online classrooms, industrial training, aircraft, maritime, military, and gaming, and highlight ethical issues, limitations, and potential pitfalls of the technology.

    \item  The paper also identifies key challenges and future research directions and make recommendations concerning the metaverse for educational services in the context of XR and IoE frameworks. 
\end{itemize}

The rest of the paper is structured as follows. Section~\ref{sec:metaedu} describes the key applications of a metaverse in the education sector. Section~\ref{sec:tech} provides a detailed overview of the technical approaches used for metaverse. Section~\ref{sec:challenges} outlines the key research challenges and future directions. Section~\ref{sec:conclusion} finally concludes the paper.

\begin{table*}
\scriptsize
\caption{Comparison of our review paper with related metaverse survey papers.} 
\label{tab:survey}
\begin{tabular}{p{1.75cm}p{0.8cm}p{1.5cm}p{0.6cm}p{0.6cm}p{0.8cm}p{1cm}p{7.0cm}}
 \hline
\textbf{Reference}
& \textbf{Year}
& \textbf{Application}
& \textbf{XR}
& \textbf{IoE}
& \textbf{Education}
& \textbf{References}
& \textbf{Main Topics}
 \\ \hline \hline
Falchuk~{et al.}\cite{falchuk2018social} & 2018 & Privacy in social metaverse & \xmark{} & \xmark{} & \xmark{} & 21 & Focuses on immersive social platforms with the privacy aspects incorporated, thereby provisions promising platforms with enhanced and persistent physical and virtual space interactions. 
\\ \hline
Nevelsteen~{et al.}\cite{nevelsteen2018virtual} & 2018 & Technologies for virtual world & \checkmark{} & \xmark{} & \xmark{} & 73 &  Focuses on the technology classification to build up a virtual world in comparison with other recent supporting technologies. 
\\ \hline
 Lee~{et al.}\cite{lee2021creators} & 2021 & Computational Arts & \checkmark{} & \xmark{} & \xmark{} & 223 & Focus on artworks by blending virtual and physical objects, to facilitate immersive arts, user-centric creation, and robotic arts through the expanded horizon of the metaverse. 
 \\ \hline
 Shen~{et al.}\cite{shen2021promote} & 2021 & User purchase promotion & \checkmark{} & \xmark{} & \xmark{} & 156 & Exploit virtual commerce considering various influential factors of customer behavior along with the synergy of design artifacts. 
 \\ \hline
Kye~{et al.}\cite{kye2021educational} & 2021 & Education & \checkmark{} & \xmark{} & \checkmark{} & 22 &  Exploit metaverse for educational services with immersive virtualization for establishing new means of social communication. 
\\ \hline
 Lee~{et al.}\cite{lee2021all} & 2021 & Development status & \checkmark{} & \xmark{} & \xmark{} & 711 & Targets six user centric features of Metaverse such as Accountability, Trust, Privacy \& Security, Creation of content, Avatar, and Virtual Economy cellular. 
 \\ \hline
 Ning~{et al.}\cite{ning2021survey} & 2021 & Development status & \checkmark{} & \xmark{} & \xmark{} & 36 & Focus on Metaverse development status from five different perspectives. 
 \\ \hline
Park~{et al.}\cite{park2022metaverse} & 2022 & Social value enhancement & \checkmark{} & \xmark{} & \xmark{} & 357 & Targets on realizing the possible components of metaverse for creating a social value by considering the implementation, application, and user interactions. 
 \\ \hline
 Chen~{et al.}\cite{chen2022exploring} & 2022 & Healthcare & \checkmark{} & \xmark{} & \xmark{} & 50 & Explores the strategies in health informatics with the laws in the application of health metaverse from the perspectives of socialization, intelligence, knowledge, and digitalization. 
 \\ \hline
 Ynag~{et al.}\cite{ynag2022fusing} & 2022 & Blockchain and AI fusion & \checkmark{} & \checkmark{} & \xmark{} & 97 & Briefly survey Blockchain, AI, and their integration with metaverse components. 
 \\ \hline
 Wang~{et al.}\cite{wang2022survey} & 2022 & Security, and privacy & \checkmark{} & \xmark{} & \xmark{} & 188 & Focuses on the security and  privacy preservation schemes, and investigated the countermeasures for  the metaverse.  
 \\ \hline
 Our survey  & 2022 & Education, Training and Skill development & \checkmark{} & \checkmark{} & \checkmark{} & 171 & Provide survey about the usage of metaverse for education, training and skill development application with the support of XR and IoE. Further, challenges, solutions and opportunities in metaverse domain are provided.
 \\ \hline

\end{tabular}
\end{table*}

\section{Metaverse Applications in Education and Training}
\label{sec:metaedu}

\begin{figure*}[!t]
\centering
\centerline{\includegraphics[height=17cm, width=14cm]{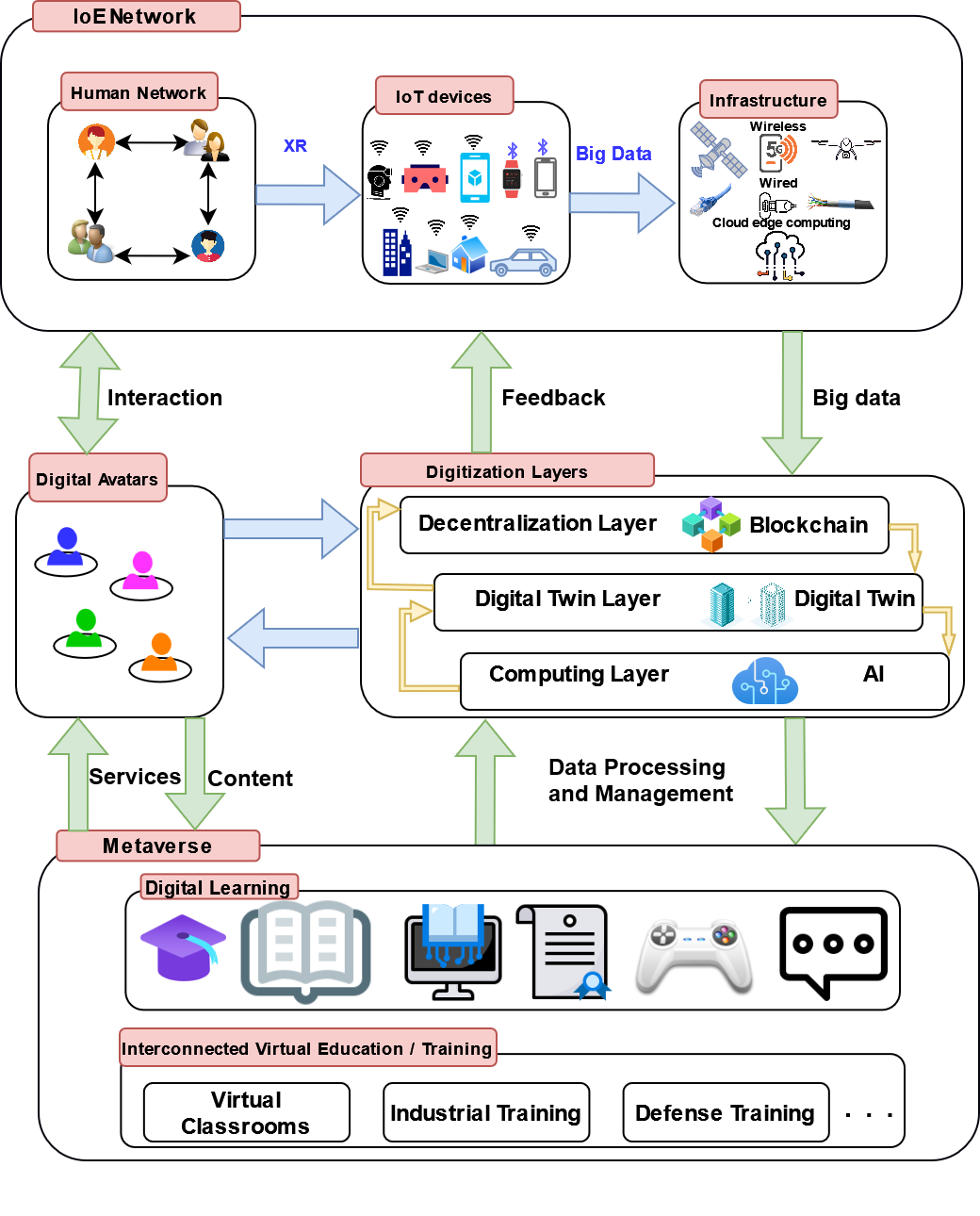}}
\caption{The Digital Interconnection among the Physical and Virtual Worlds through Metaverse for Immersive Education and Training.} 
\label{fig:Metaverse-IoE-XR}
\end{figure*}
\begin{figure}[!ht]
  \centering    
  \includegraphics[width=0.5\textwidth]{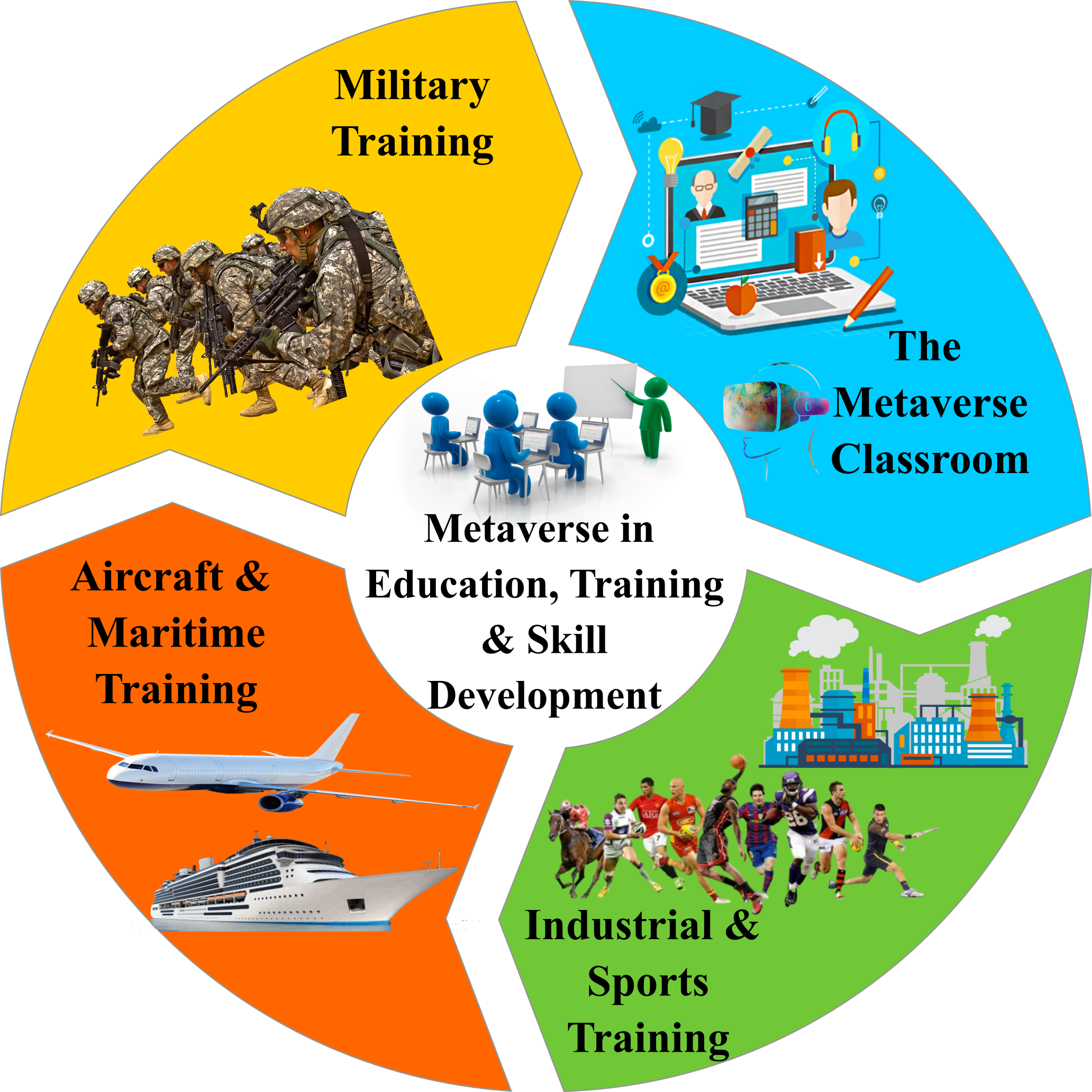}
   \caption{Few Metaverse Applications largely driven by XR and IoE}
   \label{fig:metaapp}
\end{figure}

In this section, we discuss some key aspects of education, training, and skill development that can benefit from the metaverse. The major benefits of metaverse over conventional online educational platforms can be hand-crafted by establishing digital interconnections among the physical and virtual worlds. The illustration of the relationship between the metaverse, digitization processes, and IoE networks with XR support is represented at a high level in Fig.~\ref{fig:Metaverse-IoE-XR}. The figure highlights the role of IoE, XR, and other supporting technologies to overcome the existing challenges of education and training by imparting digital interconnection among the physical and virtual worlds through the metaverse. Fig~\ref{fig:metaapp} shows few of the popular metaverse driven education, training and skill development applications driven by XR and IoT technologies.

\subsection{The Metaverse Classroom}
XR technology brings educational services closer to the students and learners in remote places. This minimizes the need to travel and emphasizes interactive and collaborative means of remote education. In~\cite{zikky2018interactive}, the authors demonstrated the efficacy of solar system learning through interactive VR services and the provision of an immersive learning environment with virtual meeting spaces along with the digital avatars of the teachers and learners.

The traditional practices in education through classroom and online platforms could be turned into virtual spaces through the metaverse and its supporting technologies. Some recent works exploit the potential of metaverse-based platforms for exploring educational, training, and learning activities through information technology services, such as cloud service, AI, and big data. This can help in shifting the focus of educational institutions towards metaverse-based education~\cite{jeon2021exploring}. In such a platform, students can gain an immersive learning experience and feel their physical presence in the collaborative classroom environment regardless of their physical location. Despite the technical and moral impediments, the metaverse imparts more interactions and cooperation among students resulting in a healthy learning experience. 

In the literature, the natural visual cognition capabilities of humans have been well exploited through the immersive metaverse platforms for making education to be more interactive and immersive across multiple ranges of cultural backgrounds~\cite{han2020visual}. Metaverse has been used as a digital education tool by including various information and communications technology (ICT) tools and technology for promoting education and learning through collaborative platform~\cite{nevelsteen2018virtual}. In \cite{nevelsteen2018virtual}, its impact was validated through the observed enthusiasm and cooperation of students beyond the classroom environment.

The metaverse can also help teachers in several ways. For instance, with the potential of metaverse implementation through AR for educators, in~\cite{maccallum2019teacher} the authors explored the potential of AR tools in further enhancing the pre-service and in-service experience of teachers. Recently, in another similar work~\cite{jeong2021teaching}, for enhancing the teaching skills of pre-service teachers, a metaverse environment is used with the support of the Virbela platform. The experimentation involved analyzing micro-teaching for a group of pre-service teachers in virtual teaching through their digital avatars. The authors analyzed the teachers' competence in handling the queries and suggestions. Such practices help teachers in developing their teaching skills by closely involving them in their profession in a more realistic and immersive way during imparting training and skill enhancement activities for the students. Furthermore, the audio metaverse through virtual mainstreaming enables the provision of interoperable experiences by rendering spatial sound along with the XR-enabled video streaming services~\cite{jot2021rendering}.  

The technology also allows the students to closely analyze different historical and architectural sites. For instance, Gaafar et al. \cite{gaafar2021metaverse} analyzed the potential of the metaverse in architectural education by providing the students and teachers with the opportunity to analyze heritage sites through virtual models of heritage buildings and their interactive 3D models. The authors concluded that being examined in the fully immersive experience through the interactive models of cultural heritages, allows the learners to observe fine details.

\subsection{Industrial Training}
Sustainable evolution of industrial infrastructure in terms of design and training of the employees is an important aspect of the modern industries especially after the fourth industrial revolution \cite{jagatheesaperumal2021duo}. Thanks to the metaverse and associated technologies, the training of employees have been also enhanced in industries allowing the workers a hands-on experience on different tasks in a risk-free environment. Moreover, AR-based training within industries helps to analyze customer demands, classify the requirements in industries, and use sustainable value proposition design strategies~\cite{werrlich2017demand}.

In the industrial assembly system powered by metaverse, we can visualize the entire factory through simulations. Global teams can collaborate to operate in synchronization using different software packages, such as CATIA, Revit, or point clouds to design and plan the factory in real-time~\cite{swikatek2018industry}. For instance, the 3D-enabled capability to operate in a perfect simulation has revolutionized the planning process of BMW car manufacturing plants. They regularly reconfigure their factories and train their employees to accommodate new vehicle launches. Planning experts located in different parts of the world could be trained and involved in testing new designs through a metaverse. 

Metaverse in industries with IoE-enabled machines and digital humans allows sharing workspace with many robots that make jobs much easier. As robots are crucial for the modern production system, they are involved in logistics for improving the material flow in the production process. This agility is highly mandatory as the volume of tasks done in each modern industry is huge and most customers demand custom-made designs for their products. The metaverse frameworks could also operate on the synthetic data, apart from the data acquired through IoE devices. From the millions of synthetic images generated across diversified categories of environments, they assist in teaching the robots~\cite{lee2021all}. Domain randomization could generate an infinite permutation of photorealistic objects, textures, orientation, and lighting conditions~\cite{assadzadeh2022vision}. It is an ideal platform for generating ground truth images and objects, whether it is meant for detection, segmentation, or depth perception for the metaverse platform. 

Moreover, virtual factory visits enabled through the metaverse could be made feasible through the digital twins. Metaverse platforms driving digital twins enable the industries to constantly innovate, collaborate, learn and train the production network toward innovations. Subsequently, this will reduce the planning times, and improves flexibility, and precision resulting in an efficient planning process. 

Similarly, the fashion industry has benefited through metaverse, by  allowing designers to make outfits that fit the customers with trails made over digital avatars and customizing the outfits according to their best fit demands. Metaverse helps to choose the outfit based on gender, age, body shape, and ethnicity. This helps the designers to learn the customer demands in an automated way and drives them to make revolutionized designs for the customers. 

\subsection{Aircraft and Maritime Training}
The potential of AR technology in the aviation, automotive, and astronautics industries has been well utilized for a long period for several tasks including environment monitoring, control, and learning purposes~\cite{macchiarella2005augmented}. Regenbrecht et al.~\cite{regenbrecht2005augmented} proposed and tested an aerospace training platform by developing augmented scenes through AR in association with state-of-the-art machine vision and computer graphics. Similarly, XR and other metaverse technologies has been also used for training workers to carry out maintenance and inspection process in the aviation industries. The literature reports the proven effectiveness of the technology in the task. For instance, Eschen et al.~\cite{eschen2018augmented} demonstrated that the deployment of XR technology for training is less error-prone and takes on less effort as well as less time-consuming. Another study reports the effectiveness of AR in operations to support, train, and maintenance of aircraft~\cite{de2010augmented}. The study of the technology results in a significant reduction in errors during the training and subsequently reduces procedure violations. These studies hinder the motivation towards modern-day metaverse to further enhance the performance of the learning platform with the significant reduction in training time and provisions of cost-effective strategies. 

The literature already reports some efforts in this direction. For instance, Siyaev et al.~\cite{siyaev2021neuro} developed a simulator for training employees to carry out efficient maintenance of Boeing-737 aircraft through metaverses along with the 3D models of the aircraft. The simulator is evaluated in terms of the amount of technical guidance and directions provided to field experts. This metaverse-based aircraft maintenance system provides a more economic and scale-able solution for aviation educational institutions. In another similar work by the authors in ~\cite{siyaev2021towards}, a CNN architecture is used for enabling the learning and classification of audio features for identification of the commands for controlling the virtual digital twin of the Boeing 737 aircraft. Further, this work has reported higher prediction accuracy with improved training capabilities through immersive interaction and effective control of virtual objects in the 3D twin model of the aircraft. More recently, Kim et al.~\cite{kim2021study} employed metaverse technology for training employees to cope with the cybersecurity risks involved in maritime industries. It includes the development of educational content for addressing cybersecurity issues through virtual ships, maritime accident reproduction, maritime operations, and related online learning content.


\subsection{Military Training}
In the defense sector, training of the troops plays an imperative part, as it is not always feasible to place the military personnel in active war locations for training. Through Distributed Interactive Simulation (DIS) protocol, earlier the Defense Advanced Research Projects Agency (DARPA) used different military simulations with the support of advanced high-level architectures for providing rich collaborative training and preparation of war strategies. XR technology could be used to replicate such an environment for training the soldiers more realistically. It can help to train the soldiers in a similar location created in a virtual environment allowing them to be prepared for dynamic adaptation. It also results in a significant reduction of expenses involved in traveling and shifting goods for troops' exercises. Moreover, the use of virtual objects helps to train the soldiers with special armed equipment without putting them at risk. However, even though, a complete replacement of military training through XR could be established with state-of-the-art techniques, the integration of IoE and metaverse technology could further enhance it. Although AR can't completely supplant conventional military preparation, there are now frameworks that kill the need to go to far-off areas and assist troopers with preparing without placing them at risk.


The diversity provided by technology compared to real world military training shifted the focus from traditional training methods to military simulations~\cite{shen2020study} by providing immersive training through the virtual environment. To this aim, several interesting solutions have been proposed. For instance, DEIMOS Military VR Trainers~\cite{nakamura2021science}, which is a metaverse-based military training, assists in creating scenarios with a diversified range of environmental conditions, particularly meant for professional military training on shooting, tactical behavior, and observation. This technology is used in the Korean military academy, which integrates spatial synchronization and precise accuracy in the shooting drill environment. In such a metaverse military drill environment, it provides real war scenarios and enables the trainers to react based on environment, precision target hitting, and provisions interactive training platform.




\subsection{Discussion and Summary of Lessons Learned}
We summarize the content of provisioning metaverse-driven educational services for various user scenarios with the key characteristics and probable solutions. In Table~\ref{tab:challenges}, we summarize key objectives/goals, challenges, and countermeasures for handling challenges associated with metaverse-based training in different application domains. From the table, it is obvious that for achieving the goal of an immersive learning platform through metaverse, the mentioned key challenges need to be addressed with appropriate countermeasures. After discussing the relevant research on metaverse for education platforms, we draw insights on providing favorable countermeasures to address those challenges: 
\begin{itemize}
    \item Multimodel medical data streamlining in healthcare education could be handled effectively through intelligent cloud services and quantum computing.
    \item With appropriate choice of XR applications and IoE devices, immersive collaboration and interactions in online education could be ensured.
    \item Skill enhancement of industrial workers for handling challenging sections could be deployed through metaverse-driven collaborative robots to achieve appropriate design, development, and supply chain management. 
    \item 3D twin models of aircraft, could assist the trainers to effectively explore the control and functional modules present in aircraft.
    \item Elimination of cybersecurtiy issues in the marine sector through metaverse ensures to provide robust mechanisms to safeguard against threats.
    \item Adverse condition anticipation in war scenes for military personals could be dealt with the quality instructions and dynamic situation handling through metaverse.
    \item Upskilling of creativity in arts and skill enhancements in gaming could be driven through metaverse with accurate object recognition and handling. 

\end{itemize}


\section{Technology Enablers for Metaverse in Education}
\label{sec:tech}

In this section, we provide an overview of technical approaches for the development of metaverse in general and in the education sector in particular. 

\begin{table*}
\scriptsize
\caption{Summary on the role of XR for metaverse-driven education and training applications.} 
\label{tab:summaryXR}
\begin{tabular}{p{2.5cm}p{3.0cm}p{4.0cm}p{7.0cm}}
 \hline
\textbf{Reference}
& \textbf{Applications}
& \textbf{XR Solutions}
& \textbf{Major Contributions}
 \\ \hline \hline
Andrews~{et al.}\cite{andrews2019extended} & Medical practice & Touch-free interface and 3D visualization & Discuss the catheter tracking, visualize scars and patient anatomy. 
\\ \hline
Doolani~{et al.}\cite{doolani2020review} & Manufacturing training & Workforce training & Introduces XR in training the workforce on maintenance and assembling tasks. 
\\ \hline
Palmas~{et al.}\cite{palmas2020defining} & Industrial training & Innovative corporate training  & Provides a set of characterisations and competencies to handle XR-enabled training
\\ \hline
Zweifach~{et al.}\cite{zweifach2019extended} & Medical education & Realistic simulation & Envision the concept of adopting XR tools in medical education. 
\\ \hline
Gandolfi~{et al.}\cite{gandolfi2021situating} & Teacher training & 360 video research &  Present the compatibility of XR by enabled training for teachers. 
\\ \hline
Stanney~{et al.}\cite{stanney2021performance} & Casualty care training &  Multi-faceted paradigm & Introduce a competency-based training platform through AI models.
\\ \hline
alizadehsalehi~{et al.}\cite{alizadehsalehi2020bim} & Construction industry & Simulate construction project & Implement a BIM-based XR process/workflow.
\\ \hline
Ilic~{et al.}\cite{ilic2021needs} & Higher education & Realistic learning-by-doing & Enhances collaborative learning skills through XR, AI and ML.
\\ \hline
Kaplan~{et al.}\cite{kaplan2021effects} & Training enhancement & meta-analysis & Enhances collaborative skills among students.
\\ \hline
Kim~{et al.}\cite{kim2021adaptation} & Nursing skill training & Smart glasses & Assists in self-practice and learning at own pace.
\\ \hline
Xi~{et al.}\cite{xi2022challenges} & Workload & Subjective workload measurement & Cost and resources reduction in operating through metaverse are assessed.
\\ \hline
Heirman~{et al.}\cite{heirman2020exploring} & Firefighting training & Mixed reality simulator & Adapting of the fire hose controller based on the scenario is trained.
\\ \hline
Mcguirt~{et al.}\cite{mcguirt2020extended} & Nutrition education & Descriptive observations & Dietary behaviors among the community are assessed. 
\\ \hline
Ong~{et al.}\cite{ong2021applications} & Ophthalmology & Ophthalmoscopy simulators & Improves procedural success and reduces complication rates in ophthalmic surgery.
\\ \hline
Logeswaran~{et al.}\cite{logeswaran2021role} & Healthcare education & Pedagogical models & Improved learning outcomes were identified through learner-centred models.
\\ \hline
Kosko~{et al.}\cite{kosko2021conceptualizing} & Teacher education & Representations of practice & Perceptual capacity assessment for theory and practice was performed.
\\ \hline
Le~{et al.}\cite{le2022narrative} & Sports training & Ecological dynamics & Motor and perceptual-cognitive skills improvements were identified.
\\ \hline
Zwolinski~{et al.}\cite{zwolinski2022extended} & Management education & Projector, mobile, HMD AR/VR & Business environment enhancement skills were imparted through different approaches.
\\ \hline
Koh~{et al.}\cite{koh2021applications} & Ophthalmology & Ophthalmic surgical simulators &  Presented proof-of-concept based on ocular imaging data.
\\ \hline
Mystakidis~{et al.}\cite{mystakidis2021motivation} & Community building & Gamify distance education & Formulates the recommendations for building practitioners. 
\\ \hline
Fast~{et al.}\cite{fast2018testing} & Manufacturing & Remote guidance & Learning, operational and disruptive phases were explored in 6 case studies.
\\ \hline
Al-Adhami~{et al.}\cite{al2019extended} & Construction quality & BIM-based XR & Quality control inspection was experimented on construction site.
\\ \hline
Parsons~{et al.}\cite{parsons2020extended} & Neurosciences & High-dimensional simulations & Enhanced ecological validity is ensured in the clinical, and social interactions.
\\ \hline
Koo~{et al.}\cite{koo2021training} & Lung cancer surgery training & Smart operating room & Reflecting on the haptic inputs of the surgeon's tactile sensations provides training.
\\ \hline
Zagury~{et al.}\cite{zagury2022current} & Otolaryngology training & Evaluation framework. & Educational outcomes and skills transfer in bone surgery were analyzed.
\\ \hline
Jervsov~{et al.}\cite{jervsov2020digital} & Control system & Digital twins  & Real-time hardware-in-the-loop (HIL) simulation for control applications.
\\ \hline
Liang~{et al.}\cite{liang2021enhancing} & Stroke assessment & Training mannequin in simulation & Traceable symptom of stroke were easily identified in the clinical training.
\\ \hline
Alnagrat~{et al.}\cite{alnagrat2021extended} & Virtual laboratories & Virtual environment of things & Established a virtual training platform to increase the efficiency of the students.
\\ \hline
Yang~{et al.}\cite{yang2020xr} & Education framework & XR-Ed framework & Increases efficiency of student education through the design spaces provisioned considering six dimensions.
\\ \hline
Andrade~{et al.}\cite{andrade2019extended} & IoT scenarios & 3D visualization & Developed a data communication model for translation of IoT data into XR scenarios, events or objects.
\\ \hline
El-Jarn~{et al.}\cite{el2020can} & Design process & Cocreation and Codesign & Establishes an unique platform to improve, co-create, and co-design the visuals in the design process.
\\ \hline
Shankhwar~{et al.}\cite{shankhwar2022interactive} & Arc welding training & 3D interactive interface & Guides with visual aids on correct welding position and procedures.
\\ \hline
Lopez~{et al.}\cite{lopez2021extended} & Medical education & Holographics & Holographic rendering and depiction of anatomical structures extends better clinical care.
\\ \hline
Guo~{et al.}\cite{guo2021development} & Educational development & Bibliometric analysis & Provides suggestions for sustainable development in the education sector.
\\ \hline
Minchev~{et al.}\cite{minchev2019extended} & Digital society & Society 5.0 & Enables foreseeing the future digital society on new cognitive skills and behavior.
\\ \hline
Goh~{et al.}\cite{goh2021virtual} & Surgical training & Immersive XR & Assists in visualizing anatomy of patients in real-time for the surgeons involved in knee arthroplasty
\\ \hline
Cross~{et al.}\cite{cross2022using} & Flight simulator & Gaming consoles & Simulator provides empirical evidence on the challenges faced due to hardware constraints by the trainers.
\\ \hline
Gandolfi~{et al.}\cite{gandolfi2020extended} & Presence scale & Multimodal presence & Improvement in the efficiency of the educators and instructors were observed along with their cognitive skills.
\\ \hline
Hoover~{et al.}\cite{hoover2021designing} & Expertise instructors  & Head mounted displays & Suggestion on design recommendations for XR training developers based on expert instructor behaviors.
\\ \hline
Pomerantz~{et al.}\cite{pomerantz2019teaching} & Teaching and learning & Exploratory evaluation & Articulates the learning experience through several learning goals and effectively meeting them.
\\ \hline
Ccoltekin~{et al.}\cite{ccoltekin2020extended} & Spatial sciences & Human factor & Identifies the role of XR in spatial science and its associated research platforms that hinders the scientific interest.
\\ \hline
Xing~{et al.}\cite{xing2021historical} & Educational trend analysis  &  Interactive connections &  Indicates the futuristic view on the education and industrial development trends through XR.
\\ \hline
Southworth~{et al.}\cite{southworth2020use} & Cardiology & 3D visualization & Provides guidance for cardiologists in terms of cardiac rehabilitation, planning and other procedures.
\\ \hline
\end{tabular}
\end{table*}

\begin{table*}
\scriptsize
\caption{Summary on the role of IoE for metaverse-driven education and training applications.} 
\label{tab:summaryIoT}
\begin{tabular}{p{2.5cm}p{3.0cm}p{4.0cm}p{7.0cm}}
 \hline
\textbf{Reference}
& \textbf{Applications}
& \textbf{IoE Solutions}
& \textbf{Major Contributions}
 \\ \hline \hline
Bandara~{et al.}\cite{bandara2016evolving} & Student performance enhancement & Global higher education & Catering to the demands of students, filters out essential educational contents with enhanced performance.
\\ \hline
Fiaidhi~{et al.}\cite{fiaidhi2019internet} & Automation & M2M & Democratization of automation skills are perceived through haptics, robotics and IoE in real-time. 
\\ \hline
Bachir~{et al.}\cite{bachir2019internet} & University 4.0 & CPS & Proposed educational ecosystem through CPS as a key component in teaching and learning process.
\\ \hline
Gul~{et al.}\cite{gul2017survey}& Education & Smart campus & Demonstrated the applications and usefulness of IoT
in the education sector.
\\ \hline
Chou~{et al.}\cite{chou2021application} & Educational institutions & Sensor devices & Focus on attendance tracking in educational institutions and ensures campus safety.
\\ \hline
Abd-Ali~{et al.}\cite{abd2020survey}& Development of education & Digital campus & Development of education envisioned through smart class and smart laboratories
\\ \hline
Al-Emran~{et al.}\cite{al2020survey}& Social education & Wearable technologies & Summarized the prospects of IoT in diversified range of educational services.
\\ \hline
Ding~{et al.}\cite{ding2020application}& Physical education & Cloud platform & Designed XR system for physical education that assists in promotion effect, and emphasizes a scientific reference.
\\ \hline
Ramlowat~{et al.}\cite{ramlowat2019exploring}& Design education & M2M & Improvement in the efficiency of teaching and learning were assessed.
\\ \hline
Burd~{et al.}\cite{burd2018courses}& Computer science education & Data analytics & Summarizes four major strategies for computer science educators by integrating IoT into their curricula.
\\ \hline
Cornetta~{et al.}\cite{cornetta2019fabrication}& STEM education & Fabrication-as-a-service & Provisions an innovative web-based master-slave approach to monitor and control the Fab lab in schools.
\\ \hline
Pervez~{et al.}\cite{pervez2018role}& Higher education & Smart lesson plans &  With specialized strategies suggests optimal teaching model with improvements.
\\ \hline
Shaikh~{et al.}\cite{shaikh2019conceptual}& Higher education & User behavior & Demonstrated a conceptual framework for education with in-depth network analysis approach.
\\ \hline
Al-Malah~{et al.}\cite{al2020enhancement}& Intelligent schools & Smart classroom & Scientific research and educational services developments were assessed. 
\\ \hline
Jean~{et al.}\cite{jean2018internet}& AI classroom & Voice assistance & Serves teaching assistants in the classroom through intelligent personal assistants.
\\ \hline
Mawgoud~{et al.}\cite{mawgoud2020security}& Higher education & Social IoT & Deals with the threats from essential domains in higher education with security solutions.
\\ \hline
Turcu~{et al.}\cite{turcu2018industrial}& Industrial education & Industry 4.0 & Transformation of industries and involved business practices were analysed in industrial setting.
\\ \hline
Leisenberg~{et al.}\cite{leisenberg2019internet}& Remote labs & Data analysis & Feasible solutions on improving the teaching strategies were analysed along with the usage of cloud platforms.
\\ \hline
Hickman~{et al.}\cite{hickman2018developing}& Education and learning & Low-cost VR & The data triangulation process provides valuable suggestions on the learning outcomes and the role of emotions.
\\ \hline
Xiao~{et al.}\cite{xiao2021analysis}& Mental health education & MQTT protocol & For the chosen mental health education architecture, combination of differential privacy policies were used for security.
\\ \hline
Somantri~{et al.}\cite{somantri2019affordable}& Industrial training & Low-cost IoT & Affordable trainer kits were developed for industrial training practices for automation.
\\ \hline
Fragou~{et al.}\cite{fragou2020exploring} & Computer science education & Ubiquitous computing & Effective learning strategies  were emphasized through digital tools, affordances and other learning approaches.
\\ \hline
Jian~{et al.}\cite{jiang2020combination}& Sports rehabilitation & Wearable sensors  &  Developed system is capable of  analyzing the ECG and EMG signals of sports personals with real-time monitoring.
\\ \hline
Mershad~{et al.}\cite{mershad2018learning}& Learning management system & Integrated learning & Presented an IoT-enhanced learning management system that integrates arts, technology and science.
\\ \hline
Qi~{et al.}\cite{qi2020university} & University education & Network education & RFID tags were used to network the teachers and students in the college education.
\\ \hline
Khan~{et al.}\cite{khan2018modern}& Higher education & Industry 4.0 & Ensures to guarantee uniform vibrant
learning skills.
\\ \hline
Kravvcik~{et al.}\cite{kravvcik2018potential}& Learning and training & Big educational data & With the novel educational tools and services, effective training and knowledge acquisition was instilled.
\\ \hline
Li~{et al.}\cite{li2022intelligent}& Visual education system & Intelligent campus & More convenient means of visual educational functionalists were highlighted with intelligent equipment that promote classroom interaction.
\\ \hline
Moreira~{et al.}\cite{moreira2018internet}& Science learning & Hypersituation environments & Programmatic contents in physical and natural sciences were made ease considering the student's reality.
\\ \hline
Zhang~{et al.}\cite{zhang2021teaching}& Entrepreneurship education & Edge computing and AI & Improved understanding of entrepreneurship and innovation education system were analyzed.
\\ \hline
Xu~{et al.}\cite{xu2022combination}& Mass education & Probability model network & Bayesian network is used to integrate university citizenship with the curriculum teaching. 
\\ \hline
Gurgu~{et al.}\cite{gurgu2019does}& Business education & Blockchain and AI & Spread awareness on the technological changes to the educational institutions from the perspective of enhancing global economy.
\\ \hline
Cui~{et al.}\cite{cui2022application}& Table tennis training & AI & The knowledge information system assists in promoting the performance of table tennis enthusiasts by assessing their skills and confidence.
\\ \hline
Li~{et al.}\cite{li2022analysis}& Music education & Smart classroom  & Assists in the enhancement of cultural literacy by building an efficient music wisdom classroom.
\\ \hline
Gkamas~{et al.}\cite{gkamas2019learning}& Learning outcomes & Data science & Focuses on the upskilling the competencies of IT workforce through a macro-level design is driven by a desktop research and a survey.
\\ \hline
Bright~{et al.}\cite{bright2021integrative}& Educate disable students & Wireless assistive technologies & Improves accessibility to technology-mediated education for the disable students.
\\ \hline
Li~{et al.}\cite{li2021application}& Basketball training & Big data & Training plans are devised through gesture recognition from videos and information fusing, which delivers effective recognition of player actions.

\\ \hline

\end{tabular}
\end{table*}

\subsection{XR Technologies for Metaverse}
The impact of XR on the metaverse has made a significant contribution toward education and skill development for workforces. The process of selecting appropriate hardware and software stacks for developing and deploying a wide range of metaverse applications is closely associated with XR technology at its core.  Besides the significant role played by IoE, XR is responsible for defining the immersive experience for users in metaverse~\cite{kim2021edge}. For example, in the context of virtual meetings, the interaction with the remote users in training and educational applications in an immersive metaverse is possible only if the end devices are deployed with the provision of XR services. Furthermore, with the support of other enabling technologies, such as cloud services, it reduces the high networking costs and enables persistent training experiences for the users~\cite{ko2021functional}. Despite expensive VR headsets, XR apps and services act as the main driving force for the present state metaverse. 

The virtual, remote, and immersive experiences via XR provided by the metaverse could largely enhance the e-commerce experiences and cost reduction in travel, and information sharing~\cite{lee2021all}. To impart adaptive and sustainable education, the transformations through XR could provide the teachers and learners with opportunities for critical thinking, better communication, collaboration, and a higher level of creativity~\cite{lee2022technology}. Such multidimensional aspects in learning provide pedagogical benefits and enhanced learning experience. Table~\ref{tab:summaryXR} summarizes the works on educational, training and upskilling driven by XR services targeted for metaverse applications.

\subsection{IoE Technologies for Metaverse}
The impact of metaverse on IoE technology will ensure enhanced remote real-world training by providing smarter and better planning for real-world items through the virtual world. As digital twins represent the software model of physical systems/assets, the role of IoE and metaverse in this arena has become imperative from the research and application perspective ~\cite{han2021dynamic}. With the advancements of IoE technology and the data streams acquired through the smart devices, they serve as a basic block for the metaverse thereby provisioning immersive interconnectivity between the real and virtual worlds.

The convergence of IoE and metaverse has already proved very effective in different application domains. The metaverse in conjunction with IoE offers a new space for people to socialize, trade, game, and even participate in music concerts, more challenges need to be resolved by the research community. By enabling IoE along with XR technology, the metaverse is also aspired to establish a new perspective of collaboration between workers, trainers, and learners to operate with much higher potential. In such situations, the IoT devices and other gadgets that can assist in processing data allow AI-powered educational content processing for metaverse at the edge. This could be incorporated in the mobile towers, in the data acquisition points, or at the IoT devices~\cite{bhattacharya2021coalition}. Furthermore, the constellation of local IoT devices and educational data acquisition sources must incorporate in coordination for application in the metaverse. Table~\ref{tab:summaryIoT} summarizes the IoE driven works on educational, training and upskilling that could be tailored for metaverse applications.

\subsection{Big Data and Predictive Analytics for Metaverse}
In the modern world, social networks, industries, the healthcare sector, and people with smart gadgets are generating large volumes of data about different aspects of life including the preferences and choices of individuals. At present, there is more demand for virtual online spaces for interaction among people from a multidimensional perspective. This allows people to immerse themselves in digital content rather than simply viewing it. All these potential applications generate a large volumes of data. The data requires real-time data analytics for predicting future outcomes in metaverse-driven businesses involved in sales, marketing, advertising, and other training applications~\cite{abdoullaev2021trans}. With the bloom of spatial computing technology that specifically relies on the processing of data from XR gadgets, there are rising concerns on handling large volumes of digital data generated from  them~\cite{egliston2021critical}. Further, the co equal collaboration among the educators and learners established through spatial computing could provide potential insights on the educational content which is one of the ambitious goals of metaverse-driven educational frameworks. 

In~\cite{kim2021studyBD}, the authors analyzed the big data generated by the digital avatars on the Zepeto platform, which enabled them to communicate with a huge community. The case study was conducted to integrate real and virtual worlds through the metaverse for analyzing news articles. This could also be extended to analyze the quality of educational content shared through metaverse, which can assist in solving associated social and legal issues. The next generation of VR headsets will also collect more user data including facial recognition and even stress level detection of the users. As the technology matures, it could also collect biometric data from individuals and impart enhanced training experiences for the users. The potential of predictive analytics in the metaverse can help in extracting significant meaningful insights from the collected data. Metaverse can also assist in transforming the data with a more user-friendly and interactive means of portraying the information~\cite{stark2022future}. With the clear benefits of predictive analytics on XR and IoE data, it is imperative to use it in the metaverse for several potential educational applications.
\begin{table*}
\scriptsize
\caption{Metaverse enabled Education, Training and Skill development Challenges \& Countermeasures.} 
\label{tab:challenges}
\begin{tabular}{p{3cm}p{3cm}p{4cm}p{6cm}}
 \hline
 \textbf{Application}
& \textbf{Challenges}
& \textbf{Goal}
& \textbf{Countermeasures}
 \\ \hline \hline
 Healthcare education & Handling multi-modal medical data and streamlining. & Innovative drive in medical education. & New directions to the healthcare education sector could be driven through the integration of metaverse with AI, VR, AR, IoMT, Web 3.0, intelligent edge, cloud services, robotics and quantum computing.\\ \hline
 Online education & Collaborative efforts and interactions. & Immersive experience for teachers and students. & Appropriate choice of XR and IoE equipment with dedicated seamless connectivity targeted for meeting the demands of the teachers and learners. \\ \hline
 Industrial training & Training robots and skill enhancement for labors & Monitor and control complex manufacturing units. & Binding the hardware and software components of metaverse in the manufacturing, supply chain, design, development, and virtual warehousing to drive the market revenue and forecast the impact of technology over the next few years and make decisions accordingly.\\ \hline
 Aircraft maintenance training & Maintenance, status monitoring and control & Intuitive and efficient control of functional modules in aircraft. & 3D twin models of aircrafts helps to read the aircraft log books and records, which includes the entries of the condition of the internal equipment, status, and intimates the requirements for the learners and users in the remote place. \\ \hline
 Marine maintenance training & Handling of cybersecurity issues & Robust defense mechanisms against treats. & Integration of metaverse with blockchain based technological trends, helps to reduce errors in maintenance tasks with secured means of handling the challenges with increased safety through alerts and notifications. \\ \hline
 Military training & Replicating the war scenes and dynamic adaptation. & Trained to face adverse conditions. & Improved productivity with clear instructions through metaverse driven equipment for dynamic handling of war situations.\\ \hline
 Arts upskilling & Managing 3D virtual objects. & Imagination and creativity to reality. & Enhanced quality and accuracy with object recognition for immersive learning with applied creativity beyond the imagination.  \\ \hline
 Gaming expertise & Integration of AI for provisioning immersive experiences. & Collaborative learning & With unified and interoperable spaces rendered through the graphics, interaction with the people and objects in the virtual worlds makes the gaming platform towards incredible heights of using the metaverse of provisioning diversified range of education, training and skill development applications. \\ \hline
\end{tabular}
\end{table*}

\subsection{Machine Intelligence}
On the road map to unleash the true potential of AI in the metaverse, machine intelligence evolves from ML by prioritizing the goals through deductive logic, where the learning algorithms fine-tune messaging, marketing, and online interactions. Machine intelligence in conjunction with the XR and IoE contributes to make a significant impact on the creativity in educational services through metaverse. Teaching through gestures interpreted through eye glances of the teacher, their emotion, and communicating cues could serve as a part of sophisticated enablers in the educational sector via metaverse. In fact, machine intelligence supports the metaverse by serving as a design advisor. Being a critical part of the service architecture, it can be integrated with low code and no code platforms~\cite{lim2022realizing}. Moreover, as the virtual world gets denser with virtual avatars, AI-based design for the fabrication of metaverse chips could also assist programmers in generating code~\cite{abdoullaev2021trans}. However, unleashing the full potential and impact of machine intelligence on this virtual landscape in the education sector demands creative progress from the developers for supporting and enhancing the features of virtual education and training platforms.

The literature reports several studies emphasizing the potential of machine intelligence in the metaverse for educational and training services. For instance, Li et al. ~\cite{li2022sadrl} emphasize the promotion of integrated machine intelligence and human experience for provisioning an elegant framework that could drive the metaverse with a rich experience for the learners involved. 

\subsection{Blockchain Adoption}
Blockchain adoption grows exponentially across various industries \cite{hewa2021survey}. It is a distributed ledger technology that enables decentralized and autonomous management of assets and applications. Despite the popularity and benefits of the metaverse, securing the data and digital information of its users is a common concern for learners and educators. Blockchain could provide a promising solution in this arena, due to its unique characteristics of decentralization, immutability, and transparency. Its support in metaverse applications is enabled through the programmability of blockchain, which could drive the educational content by putting together smart contracts among decentralized locations~\cite{sanka2021survey}. 

The literature reports several interesting solutions for the integration of blockchain in metaverse applications. For instance, Gadekallu et al. ~\cite{gadekallu2022blockchain} provide a detailed overview of the approaches to seamless integration of blockchain in the metaverse by considering the data acquisition, storage, privacy preservation, and interoperability aspects. Furthermore, the impact of metaverse on its supporting technologies, such as IoE, digital twins, AI, and big data were explored in association with the role of blockchain in securing the platform. This stands as a crucial drive towards the implementation of secure educational content for the metaverse-based educational frameworks.

Trust and authority enhancement through blockchain adoption for the metaverse enables more IoT devices to participate and assist in developing educational content from active gaming to embedding financial services with the contents~\cite{kommadi2022blockchain}. However, as the social scalability of metaverse-enabled educational services may take a longer period, trustless education contents, components and contracts may be rooted up. We could gain benefits from earlier adoption of blockchain to establish on-chain educational data feeds along with the bloom of virtual commodities~\cite{tan2021ethical}. Furthermore, the avatar customization and new generation educational feed could be made more immersive and trustworthy in the metaverse through the widespread adoption of blockchain and by associating them through cloud computing components. 


\subsection{Low-Code Platforms}
In recent years, the use of low-code and no-code application platforms (LCAP) has replaced hand-coding of processes and accelerated higher-level abstractions in XR-based applications. It enables even the non-programmers to perform several tasks generally needing programming expertise. According to Gartner, most of the leading industries have started using LCAPs to operate at least part of their infrastructure~\cite{bock2021low}. This trend can help the developers to focus more on the complexity and nuances of developing XR-based educational services in metaverse instead of worrying about the complexity of developing and deploying Internet-scale applications. Furthermore, a diversified set of creator tools are available to develop metaverse contents with sophisticated features and for provisioning better support for commerce and education sectors~\cite{sanchis2020low}. Although it is challenging for businesses to scale down the technology to individual educational institutions, there are numerous occasions when IoE enabled devices in the hands of individuals, could drive towards the most cost-effective option for the enterprises.

As the software industries frequently impart vital technical changes, low-code platforms highlight a high level of concern for data insights and provision of efficient and reliable solutions~\cite{laato2022trends}. Furthermore, the use of cloud services as the dominant computing platform aids in creating rich immersive experiences for learners in the metaverse environment. The metaverse along with XR will gradually be built by a larger number of producers, with a wider catalog of plug-in apps and business logic to support them. However, the present literature on LCAP does not fully represent the impact or the reasons behind the educational sector's use of these XR and IoE platforms~\cite{ihirwe2020low}. With multiple data endpoints, a larger support from LCAP could be accounted for in virtual educational platforms and it ensures scalable and secure solutions.

\subsection{Accelerating Distributed Networks}
The computing networks are said to be distributed when the programming and the data are spread out across more than one computing resource. Accelerating the performance of distributed networks in terms of its networking speed, latency and concurrency could be driven through 5G and 6G services~\cite{swikatek2018industry}. This acceleration will sustain the metaverse, and also helps to provide more fascinating applications in the education sector by enabling the users to share real-time educational content. As the data intensive networks are key resources in metaverse applications, its association with the distributed networks needs to be tightly coupled and accelerated~\cite{cai2022compute}. Furthermore, the metaverse frameworks need to be end-to-end optimized and the computation and communication should be promising for imparting rich educational experience. As the metaverse educational platforms are integrated with IoE, they could gain maximum benefit through edge computing. However for low performance IoE devices, it remains a huge challenge. ScissionLite~\cite{ahn2021scissionlite}, a holistic framework was developed by the authors to assist in accelerating the distributed DNN through insertion of a traffic-aware layer, thereby increasing the network traffic. This framework greatly enhanced the interference latency multifold in comparison with the conventional slicing approaches used in the learning models. Further, the interference latency could be considerably reduced through the implications of acceleration deep neural networks.

Towards the realization of edge-enabled metaverse, the authors in~\cite{xu2022full} surveyed the networking and communication aspects in the implementation of edge solutions. With distributed edge computing solutions the resource constraint challenges existing such platforms could be effectively addressed~\cite{lee2022resource}. Further, acceleration in computing enables to render an immersive experience for the learners delved into the metaverse platforms. Moreover, it also authorizes sharing of virtual user-generated educational content with low cost of data transfer frameworks. Such a decentralized, tamper-proof solution for the edge-enabled educational services in a metaverse framework enables transparent information exchange between the teachers and learners.



\section{Challenges and Issues Related to Metaverse}
\label{sec:challenges}
Despite the significant efforts for addressing various aspects of the metaverse in the field of education, its essential requirements and perspectives need the attention of the research community. This section provides a roadmap of challenges that could impact the implementation and deployment of metaverse for educational services. We also discuss the adaptable environments and consider the use of emergent paradigms from social and technological trends that could shape the future trends of the metaverse for educational services.

\subsection{Virtual Mainstreaming}
XR and particularly IoE frameworks have to deal with the genuine aspects of the virtual world to be just like the real one, as well as the trust aspects to the functioning of relationships and organizations. Establishing standards for such technology is the foundation for how we establish and assess our connections and continue to function in harmony with the legal systems. Each of these technologies is scalable because of trust. The scalability of the metaverse in conjunction with the IoE and XR technology will improve trust in the virtual domain as the trend grows with virtual objects, crypto assets, smart contracts, virtual friends and live online educational experiences. 

Since users believe that the possessions of virtual connections and data exchanges are genuine, online bullying, abuse, cheating, and cheating in relationships will all become more damaging. These issues need to be solved through proper education, training, virtual literacy, and supportive communities. However, as people respect the virtual world more, there are usually groups of people that exploit the trend for selfish benefits. Researchers aim at providing infrastructure through virtual streaming of resources and thereby ensuring to address the trust issues in each of the core domains to combat cybercrimes, such as phishing, online frauds, virus distribution, and ransomware attacks~\cite{ynag2022fusing}. 

\subsection{Challenges by Open Platforms}
Open platform systems have become a common social phenomenon mainly due to the large range of collaborators involved in developing software modules~\cite{liu2018enriching}. The graphical and immersive experiences gained through metaverse with the support of XR and IoE can be distributed outside of app stores with the open platforms. Data-oriented technology stack platforms leveraging these features help to deliver efficient metaverse experiences, particularly for the educational sector. However, the exponential growth in the usage of open source apps in permission-less social networks, operating systems, and PCs often impart challenges for the developers. Furthermore, new technology and open standards are emerging that have the potential to democratize the future of metaverse~\cite{mcveigh2019shaping}. The technology with decentralized digital identities, such as the avatars of individuals and zero-knowledge proofs must help to claim control over their valuable data. More breakthrough opportunities persist in the development of free applications with XR and IoE framework support through open platforms for the massive increase in the impact of metaverse on educational services. 

\subsection{Walled Garden Ecosystems}
The Walled Garden is the most prominent strategy used by the technological giants, such as Google, Facebook, and Amazon for restricting the user navigation within their network premises by providing access and provisioning only required operational services~\cite{key2020information}. Technological trends that impact the metaverse favor the usage of a well-organized and attractive walled garden ecosystem. The challenges that exist in the open systems are could be addressed through the walled garden, which allows the creators to collaborate, modify and link with their designs~\cite{sungurouglu2020ecological}. However, technologies for portable avatars, portable social networks, and interoperability are on the horizon, and this could help connect diverse walled gardens using open platforms while also allowing for discovery and curation options. Furthermore, there are very few walled gardens and it is required to be configured by creators and could assist in inviting other creators and joining with them to manipulate the shared components, link with them, and collaborate in the metaverse-based educational services. As the educational services have access to a vast range of audiences, the equivalent virtual world hyperlinks on websites for the metaverse and hypermedia-like structures with portals may provide immersive experiences for the users. Most end-users, service providers, and educational institutions providing metaverse-based services for the learners will also feel safe with these technological advances.

\subsection{Simulating Reality}
The notion of simulating the universe and reality is not new. It could be traced from numerous philosophies, the distinction between reality and dream, which was once termed as an illusion. With the proper usage of physics of light energy, the ray-tracing simulation defines the appearance of images and enables 3D graphics for real-time visuals~\cite{mchaney2018immersive}. The utilization of pre-rendered contents and realistic graphics for metaverse demands more computing power for the IoT devices used for such tasks. However, real-time raytracing could solve these issues by enhanced AI models that could mimic the machines and provide an immersive experience for the metaverse through virtual objects, people, places, and machines.  For instance, the data accumulated from digital twins, IoT devices monitoring traffic, and geospatial data could impose real-time data of processes, objects, machines, and people and provide real-time visualization of the data through AI and predictive analysis. Furthermore, these features add an additional layer of immersive experience for the metaverse and could assist the next-generation education with more interactiveness and engage the learners and educators with a better platform. 

\subsection{Ethical Considerations and Potential Pitfalls}

\begin{figure}[!ht]
  \centering    
  \includegraphics[width=0.5\textwidth]{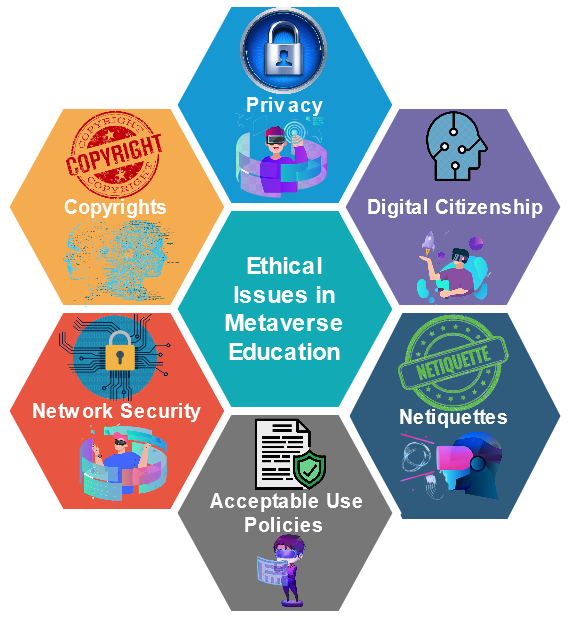}
   \caption{Salient Ethical Issues in Metaverse-based Education.}
   \label{fig:taxonomyethics}
\end{figure}

In this section, we summarize ethical considerations and potential pitfalls of using metaverse for educational services. Furthermore, we will survey the state of the current literature on the legal issues involved in the use of metaverse for educational services. Interoperability and openness of the metaverse raise the question of the ethics and values for the future Internet. The ethical judgments based on the philosophical nature of the technology imparted in the metaverse, need to identify the responsible members building this infrastructure. With the core objective of preserving human wellbeing, the ethical guidelines provided in \cite{shahriari2017ieee}  could well suit the metaverse applications with an increased focus on the use of AI for autonomous frameworks. Fig~\ref{fig:taxonomyethics} shows the taxonomy of ethical issues involved in the metaverse-based educational services.

\begin{itemize}
    \item \textit{Privacy Policies}: The XR devices along with brain-computer interfaces (BCI) could provide access to the metaverse applications, which are capable of detecting the thought process of the users. Despite using them for predicting consumer behavior, they gather private user data and keep storing them in the blockchain forever. Privacy policies of metaverse applications should be read and accepted by understanding the consequences of data leaks, misuse of data, and hacking of personal data. The authors in \cite{casey2019immersive} demonstrated potential solutions for securing such systems. For social VR learning environments, security and privacy attack trees are demonstrated in \cite{valluripally2020attack}, in which stochastic timed automata representations are used for statistical model checking. Furthermore, integrity loss and privacy leakage are also assessed before and after the adoption of the aforementioned design principles.

    \item \textit{Digital Citizenship}: Digital citizenship refers to the way of being and acting online in a safe and meaningful way. It drives towards critical thinking and not trusting everything the user observes. It ensures being safe with the information and getting connected to the intended persons. Digital citizenship ensures the metaverse users to act with responsibility in communication and behavior. The authors in~\cite{kim2018development} proposed a five factor digital citizenship model that ensures social engagement online through reasonable activities. Such models driven by the educators for the metaverse could provide reliable education and training contents as a digital citizenship scale for the learners.
    
    \item \textit{Netiquettes}: Digital etiquette is normally termed as Netiquette, where having proper communication is key in terms of asking, and responding with respect and protection. By proper greeting, asking specific queries, with professional and proper ending  Netiquette could be ensured in communication. In the metaverse-based education implemented in high schools, college or industries, being a responsible digital citizen is a key. In~\cite{soler2021netiquette}, the authors reviewed the necessity of Netiquette in online educational platforms, where information and communication technologies (ICT) have changed the way of communication and socialization. The metaverse with new reality for education and training the learners with proper Netiquette, arrests cyber-bullying, and digital scams, in the social networks.
    
    \item \textit{Acceptable Use Policies}: It refers to the set of rules or policies applied by the owner, creator or administrator of the educational content or the system, that restricts the way the content could be used. It is an integral part of the framework of security policies, where new users gain access to the resources by signing the policy. Theoretical and historical understanding of the policies on educational contents are evaluated in~\cite{brown2021steering} for steering the distance education. An online consent maturity model developed in~\cite{rooney2018online}, with a socio-technical context, argued the legalistic adoption of ethics in the paradigm for informed consent in online education. The stakeholders of educational institutions may request the learners and teachers to sign an acceptable use policy, before allocating them a network ID for gaining access to the metaverse-enabled educational contents.
    
    \item \textit{Network Security}: As security concerns are raising in immersive educational platforms, the attacks on the Head-mounted display (HMD) devices could enable the attackers to disorient users and overlay or modify the images in their vision. In \cite{casey2019immersive}, the authors demonstrated the exploitation of VR systems for controlling the users immersed in their metaverse world without their knowledge. These attacks were termed Human Joystick attacks (HJA). 
    
    \item \textit{Copyrights}: With the widespread bloom of numerous online educational platforms, the management of digital rights on the educational content against infringement of copyrights of learning resources are challenging. Blockchain-enabled digital copyright management system in \cite{rooksby2019copyrights}, ensures sharing of multimedia educational contents through private and public blockchains. It ensures data protection of educational contents, that could be appropriate choice for metaverse platforms as well. Further, in  \cite{guo2020blockchain}, the authors presented on the usage of blockchain for claiming and protection of the intellectual property rights on the ownership of technology transfer resources.
    
\end{itemize}    

As stated in the aforementioned sections, XR is all set to modify our perceptions. Such XR solutions are created in such a manner that we mistake a virtual version of the actual world for a genuine object and become engaged in it. This technology, like any other, has some negative consequences as well. It can readily create virtual false places and fake circumstances, and can also get users into mind-altering states. This can be misused to brainwash people and lead them to pursue unintended goals. As XR becomes a more powerful component of the metaverse, we still need the technology to tell what is authentic. The industry needs to invest in developing standards and policies that allow XR gadgets to operate ethically, and prevent the misuse of metaverse technology.

\section{Future Research Directions and Open Issues}
\label{sec:future}
The analysis of the metaverse in a particular application is based on multi-domain design exploration with many dimensions. This section describes the takeaways from this survey with a summarization of the relationship between the methodologies, architecture, and frameworks of metaverse used for education, training, and skill development applications. 

Most educational and training applications require a compromise between high-quality content and an immersive experience for the users. High bandwidth communication and throughput performance may not be a crucial metric for these systems. The frameworks employed in these applications should maximize the performance and consider other constraints associated with the IoE and XR devices such as power, memory, size, and other resource constraint modules for the trainers and learners. In metaverse-enabled classrooms, these constraints have a larger impact as XR and IoT devices are much more influential and largely used by the learners and trainers. Therefore, in these applications, immersive experience is prioritized with high throughput and high-speed communication. 

In this section, we discuss a range of open topics in the metaverse and beyond for future research in this domain.


\begin{itemize}

    \item \textit{Accuracy and Authentication of Material}: 
    Offering truly quality and immersive education through metaverse demands the provision of high-quality content. The usage of substandard and highly incompatible XR and IoT devices also impacts the accuracy and authentication of the educational contents. There is a need to put efforts to define a common authentication mechanism along with the features to deliver accurate and high-quality educational content. However, this initiative to create a robust educational platform needs to enable cross-border interoperability across a diversified range of supporting technologies.
    
    \vspace{1mm}
    \item \textit{Access to Reliable Broadband}: 
    Seamless Internet connectivity for metaverse-driven education largely depends on access to reliable broadband services. The larger coverage and high-speed access to educational resources make it a worthwhile option for streaming educational content. Optical communication and 5G/6G services have already taken the lead to lay the platform for ensuring high capacity with features, such as reliable video streaming and sharing of high-quality immersive educational content~\cite{zhang2022interpreting}. However, given the access to broadband services, using heterogeneous radio technologies based on their availability often affects the trust in the communication services rendered for the educational services~\cite{das2022xhac}.
    
    \vspace{1mm}
    \item \textit{Marginalized Communities}: The research community should ensure focusing on the technology and design that could satisfy the demands of the marginal group of individuals and communities participating in the metaverse-driven educational services.  This will allow them to have full access to the services provided by metaverse. Education for physically impaired learners and the people with cognitive impairments should be considered by the metaverse platform and enable fairness policy to treat such individuals~\cite{strumke2022explainability}.  Attribution-based learning models with the Shapley values, largely used in explainable artificial intelligence (XAI)  frameworks, that depend on the game theory approach to interpret explanations from the deep neural networks. This enables the marginalized community to provide immersive education~\cite{bhargava2022explainable}. Further, such XAI models are largely used in identification of vulnerable groups and aids in exploring how the decisions were made by the models. Particularly for the educational services driven through metaverse, the threats over the educational contents and the inherent contextual exploration needs attention from the research community.
    
    \vspace{1mm}
    \item \textit{Integration of mobile technologies and appropriate pedagogical methodologies}: 
    A key component in provisioning metaverse-driven educational services is its integration with mobile technologies. With a lot of pedagogical methodologies coming up with interactive mobile apps, the adoption of metaverse further enhances the usage of these apps for streamlining the education towards immersive and interactive means. Mobile technologies are important because, although people are at different geographical locations, they are interconnected through their sophisticated gadgets, particularly with IoE and XR modules. Moreover, there are provisions for imposing user-defined sharing and restrictions on educational content through the apps. However, constraints on power management, high-speed networking, and coverage limitations need to be addressed to meet the balance with the demands of using the immersive pedagogical methodologies. The influence of perceived trust and pedagogical faith on the trainers and educators are largely enhanced through the acceptance of AI-based tools and services~\cite{choi2022influence}. Encouragement on the acceptance of the education XAI platforms with metaverse could enable the teachers to make pedagogical decisions and action on the learners. With mobile technology, XR's next-generation human-centric services could be triggered through discrete event simulations driven through metaverse platforms~\cite{turner2022next,ahmed2022artificial}. This in turn, largely assists educational services with the usage of immersive XR apps and low cost VR/AR gadgets for learner-centric sophisticated educational services. 
    
    \vspace{1mm}
    \item \textit{Integrated Learning}: 
    Conventional online learning strategies with a lot of distractions are often considered inefficient. Since the gadgets must handle high-speed educational content, the capabilities of networks, smartphones, and desktops severely degrade for provisioning immersive learning strategies. Integrated learning through XR and IoE modules with the support of metaverse technology enables the integration of the educational contents through games and storylines~\cite{tsapara2022board}. However, the complete potential of a metaverse in integrated learning could be reaped with sophisticated enabling technologies and high-speed communication by addressing the challenges such as the size of devices, power management, and resource management across the integrated learning platform.
    
    \vspace{1mm}
    \item \textit{Developing Values of Responsible Usage}:
    While metaverse along with XR and IoE architecture can assist with immersive means of educational services, the users are still struggling with effective management of responsible usage of the technology. Despite having secure communication support, the physical characteristics of the IoT devices and VR/AR gadgets are often being misused without considering their values~\cite{kozinets2022immersive}. There are inimitable properties that are unique to every technology that supports the growth of metaverse. However, apart from emphasizing stringent security and privacy through robust encryption schemes, responsible usage of the technology by the end users and developers must be supported with their values. From the perspective of using the metaverse and its supporting technology for educational services, the learners, and teachers should be responsible in imparting quality educational services through the metaverse~\cite{xi2022challenges}. Even though the technological implications at present are minimal in this context, the role educational service providers are also a predominant one and their challenges need to be addressed by the research community on high priority.
\end{itemize}
\section{Conclusions}
\label{sec:conclusion}

The current generation of XR applications taught us the way to blend the physical and virtual worlds through IoE devices. Metaverse platforms incorporating the XR and IoE technologies have drawn attention in recent years from the commercial and research perspective. With these technologies promising positive impact on human lives through IoE use cases. Further, metaverse also constitutes an interactive visualization and interpretation for educational services with the chain of raw data generated from IoE devices. 
These IoE data are analyzed by digital models to provide a high level of abstraction and insight into the data streamed from IoT devices. In this survey, we reviewed the XR solutions and associated challenges in education, training and skill enhancement in various fields through IoE applications from the metaverse perspective. Specifically, we highlighted the key concepts involved in XR and IoE technology as well as the vital features incorporated for training and skill enhancement through metaverse. We also presented the contributions of metaverse using XR and IoE technology, which is used in the context of a diversified range of training and educational services. We also identified the open issues for sustaining the development of XR-based training supported through IoE devices and thereby extending its services towards the deployment of immersive education, which was beyond the scope of this survey. This article will help the readers to understand the state-of-the-art features of XR and IoE devices and their capabilities, which meet the demands of a metaverse in training, and further this paper also lists the future research prospects covering some promising directions in this field. 
\printcredits
\bibliographystyle{unsrtnat}

\bibliography{references}
\vskip3pt
\bio{}
\textbf{Senthil Kumar Jagatheesaperumal}
received his B.E. degree in Electronics and Communication Engineering from Madurai Kamaraj University, Tamilnadu, India in 2003. He received his Post Graduation degree in Communication Systems from Anna University, Chennai, in 2005. He received Ph.D. degree in Embedded Control Systems and Robotics from Anna University, Chennai in 2017. He has 14 years of teaching experience and currently working as an Associate Professor in the Department of Electronics and Communication Engineering, Mepco Schlenk Engineering College, Sivakasi, Tamilnadu. He received two funded research projects from National Instruments, USA each worth USD 50,000 during the years 2015 and 2016. He also received another funded research project from IITM-RUTAG during 2017 worth Rs.3.97 Lakhs. His area of research includes Robotics, Internet of Things, Embedded Systems and Wireless Communication. During his 14 years of teaching he has published 20 papers in International Journals and more than 25 papers in conferences. He is a Life Member of IETE and ISTE.
\endbio

\bio{}
\textbf{Kashif Ahmad} received the bachelor's and master's degrees from the University of Engineering and Technology, Peshawar, Pakistan, in 2010 and 2013, respectively, and the Ph.D. degree from the University of Trento, Italy, in 2017. He worked with the Multimedia Laboratory in DISI, University of Trento. He is currently working as a lecturer in the Department of Computer Science, Munster Technological University, Cork, Ireland. He has also worked as a Postdoctoral Researcher at Hamad Bin Khalifa University Doha, Qatar and ADAPT Centre, Trinity College, Dublin, Ireland. He has authored and coauthored more than 70 journal and conference publications. His research interests include multimedia analysis, computer vision, ML, and signal processing applications in smart cities. He is a program committee member of multiple international conferences, including CBMI, ICIP, and MMSys.
\endbio

\bio{}
\textbf{Al-Fuqaha} [S'00-M'04-SM'09] received Ph.D. degree in Computer Engineering and Networking from the University of Missouri-Kansas City, Kansas City, MO, USA, in 2004. He is currently a professor at Hamad Bin Khalifa University (HBKU). His research interests include the use of machine learning in general and deep learning in particular in support of the data-driven and self-driven management of large-scale deployments of IoT and smart city infrastructure and services, Wireless Vehicular Networks (VANETs), cooperation and spectrum access etiquette in cognitive radio networks, and management and planning of software defined networks (SDN). He is a senior member of the IEEE and an ABET Program Evaluator (PEV). He serves on editorial boards of multiple journals including IEEE Communications Letter and IEEE Network Magazine. He also served as chair, co-chair, and technical program committee member of multiple international conferences including IEEE VTC, IEEE Globecom, IEEE ICC, and IWCMC.
\endbio

\bio{}
\textbf{Junaid Qadir} [SM'14] is a Professor of Computer Engineering at the Qatar University in Doha, Qatar. He is also affiliated with the Information Technology University (ITU) of Punjab in Lahore, Pakistan. He directs the IHSAN Research Lab. His primary research interests are in the areas of computer systems and networking, applied machine learning, using ICT for development (ICT4D); human-beneficial artificial intelligence; ethics of technology, artificial intelligence, and data science; and engineering education. He has published more than 150 peer-reviewed articles at various high-quality research venues including publications at top international research journals including IEEE Communication Magazine, IEEE Journal on Selected Areas in Communication (JSAC), IEEE Communications Surveys and Tutorials (CST), and IEEE Transactions on Mobile Computing (TMC). He was awarded the highest national teaching award in Pakistan---the higher education commission's (HEC) best university teacher award—for the year 2012-2013. He has obtained research grants from Facebook Research, Qatar National Research Fund, and the HEC, Pakistan. He has been appointed as ACM Distinguished Speaker for a three-year term starting from 2020. He is a senior member of IEEE and ACM.
\endbio

\end{document}